\documentclass{pnastwo}

\usepackage[dvips]{graphicx}
\usepackage{times}
\usepackage{hyperref}

\begin{document}



%
\title{Toolbox model of evolution of prokaryotic metabolic networks and
their regulation}
\author{Sergei Maslov\affil{1}{Department of Condensed Matter Physics
and Materials Science, Brookhaven National Laboratory,
Upton, New York 11973, USA},
\thanks{Author to whom correspondence should be addressed, E-mail: maslov@bnl.gov}
Sandeep Krishna\affil{2}{Niels Bohr Institute, Blegdamsvej 17,
DK-2100, Copenhagen, Denmark},
Tin Yau Pang\affil{3}{Department of Physics and Astronomy,
SUNY Stony Brook, Stony Brook, NY 11794-3800}\affil{1}{},
\and Kim Sneppen\affil{2}{}}
%

\contributor{Submitted to Proceedings of the National Academy of Sciences
of the United States of America}
\maketitle

\begin{article}
\begin{abstract}
It has been reported that the number of
transcription factors encoded in prokaryotic genomes
scales approximately quadratically with their total
number of genes.
We propose a conceptual explanation of this finding
and illustrate it using a simple model in
which metabolic and regulatory networks of prokaryotes
are shaped by horizontal gene transfer of
co-regulated metabolic pathways.
Adapting to a new environmental condition monitored by a new
transcription factor (e.g. learning to utilize another
nutrient) involves both acquiring new enzymes as well as
reusing some of the enzymes already
encoded in the genome. As the repertoire of enzymes
of an organism (its toolbox)
grows larger, it can reuse its enzyme tools more often, and
thus needs to get fewer new ones to master each new
task. From this
observation
it logically follows that
the number of functional tasks and their regulators increases faster than
linearly with the total number of genes encoding enzymes.
Genomes can also shrink
e.g. due to a loss of a nutrient from the
environment followed by deletion of its regulator
and all enzymes that become redundant.
We propose several
simple models of network evolution
elaborating on this toolbox argument and
reproducing the empirically observed
quadratic scaling.
The distribution of
lengths of co-regulated pathways in our model
quantitatively agrees with
that of the real-life
metabolic network of {\it E. coli}.
Furthermore, our model provides a qualitative
explanation for broad distributions of regulon sizes
in prokaryotes.
%
%
%
\end{abstract}

\keywords{Horizontal Gene Transfer,  Transcriptional
regulatory networks, Functional genome analysis}

\abbreviations{HGT, Horizontal Gene Transfer; KEGG, Kyoto Encyclopedia of Genes
and Genomes; TF, Transcription Factor}

\section{Introduction}

Biological functioning of a living cell involves
coordinated activity of its metabolic and
regulatory networks. While the metabolic network
specifies which biochemical reactions the cell is in
principle able to carry out, its actual operation in a
given environment is orchestrated by the transcription regulatory
network through up- or down-regulation of enzyme levels.
A large size of the interface between these two networks in
prokaryotes is indicated by the fact that nearly half
of transcription factors in {\it E.coli} have a
binding site for a small molecule \cite{babu2003},
which implicates them \cite{ananth} as potential
regulators of metabolic pathways. This interface
is further increased when one takes into account
two component systems whose sensors bind to small
molecules and only then activate a dedicated transcription factor.
Thus, at least in prokaryotes, regulation of metabolism
occupies the
majority
of all transcription factors.

Two recent empirical observations shed additional
light on evolutionary processes shaping these two
networks:
\begin{itemize}
\item The number of transcriptional regulators is shown to
grow faster than linearly
\cite{stover,nimwegen,lorenzo,tiedje}
(approximately quadratically \cite{nimwegen})
with the total number of proteins
encoded in a prokaryotic genome.
\item
The distribution of sizes of co-regulated pathways (regulons),
which in network language correspond to
out-degrees of transcription factors in the regulatory
network, has long tails \cite{thieffry}.
As a result the set of transcription factors of each organism
includes few global (``hub'') regulators controlling hundreds of genes,
many  local regulators controlling several targets each, and all regulon sizes in-between these two extremes.
\end{itemize}

A simple evolutionary model
explains both these empirical observations in a
unified framework based on modular functional design
of prokaryotic metabolic networks and their regulation.

\subsection{A toolbox view of metabolic networks.}
Metabolic networks are composed of many
semi-autonomous functional modules
corresponding to traditional
metabolic pathways \cite{kegg} or their subunits
\cite{gelfand_mirny2006}).
Constituent genes of such evolutionary modules tend to
co-occur
(be either all present or all absent) in
genomes \cite{vonmering2003sdp,gelfand_mirny2006}.
These pathways overlap with each other to form
branched, interconnected metabolic networks. Many of
these pathways/branches include a dedicated
transcription factor turning them on under appropriate
environmental conditions.
%
In prokaryotic organisms there is a strong positive
correlation between the number of protein-coding genes
in their genomes, the number of
metabolic pathways formed by these genes, the
number of transcription factors regulating these
pathways, and, finally, the number of environments or
conditions that organism is adapted to live in.

We propose to view the repertoire of metabolic
enzymes
of an organism
as its toolbox. Each
metabolic pathway is then a collection of tools
(enzymes), which enables
the organism
to utilize a particular metabolite
by progressively breaking it down
to simpler components, or, alternatively, to
synthesize a more complex metabolite from simpler
ingredients.
Adapting to a new environmental condition
e.g. learning to metabolize a new nutrient,
involves acquiring some new tools as
well as reusing some of the tools/enzymes that are
already encoded in the genome. From this
analogy it is clear that as the toolbox of an organism grows
larger, on average, it needs to acquire fewer and fewer
new tools to
master each new metabolic task.
This is because the
larger is the toolbox the more likely it is
to already contain some of the tools necessary
for the new function.
Therefore, the number of proteins encoded in
organism's genome (i.e. the size of its toolbox) is
expected to increase {\it slower than linearly} with
the number of metabolic tasks it can
accomplish.
Or, conversely, the number of
nutrients an organism can utilize via distinct
metabolic pathways is expected to scale {\it faster than
linearly} with its number of enzymes or reactions in
its metabolic network.
This last prediction is
empirically confirmed
by the data in the KEGG
database \cite{kegg}: as shown in
Fig. S6 in supplementary materials
the best powerlaw fit to
the number of metabolic pathways
vs the
number of metabolic reactions in prokaryotic genomes
has the exponent $2.2 \pm 0.2$.
This is in agreement with quadratic
scaling of the number of transcription factors
\cite{nimwegen} if one assumes that most of
these pathways are regulated by a dedicated
transcription factor.

%
\section{Results}
\subsection{Evolution of networks by random
removal and addition of pathways}

We propose a simple model of evolution
of metabolic and regulatory networks based on this
toolbox viewpoint. The metabolic network of a given organism
constitutes a subset of the
``universal biochemistry'' network, formed by the union of
all metabolites and metabolic reactions
taking place in any organism. An approximation to
this universal biochemistry can be
obtained by combining
all currently known metabolic reactions
in the KEGG database \cite{kegg}.
%
For prokaryotes, entire metabolic pathways from this universal
network could be added all at once by the virtue of Horizontal Gene
Transfer (HGT), which according
to Ref. \cite{horiz_transfer} is the dominant form of
evolution of bacterial metabolic networks.
%
%
%
%
%
Recent studies \cite{beiko} reported a number of
HGT ``highways'' or preferential  directions of
horizontal gene transfer between major divisions of
prokaryotes. As a result of
these and other constraints
the effective size of the universal network
from which an organism gets most
new pathways is likely to
deviate from
the simple union of reactions in all organisms.
Metabolic networks can also shrink due to removal of
pathways. This often happens when a nutrient disappears from the
environment of an organism over an evolutionary significant
time interval (see ``use it or loose it" principle by
Savageau \cite{savageau}). A massive elimination of pathways
occurs e.g  when an organism becomes
obligate parasite fully relying on its host for
``pre-processing'' of most nutrients.

The state-of-the-art information on metabolic networks
is not adequate for a fully realistic modeling of
their evolution.
Fortunately, faster-than-linear scaling of the
number of pathways and their regulators with the number of genes
is the robust outcome of the toolbox evolution scenario and
as such it is not particularly sensitive to topological structure
of the universal biochemistry network.
%
In particular we found (see Fig. S1) essentially identical scaling in two models
using two very different variants of the universal
biochemistry network:
\begin{itemize}
\item
the union of KEGG reactions \cite{kegg} in all organisms. The part of
this network connected to the biomass production consists of
$N_{univ} \simeq 1800$ metabolites;
%
%
\item
a random spanning tree on the fully connected
graph of $N_{univ}$ metabolites.
While certainly
not realistic, this version is mathematically tractable.
\end{itemize}
Furthermore, it turned out that many other details of pathway acquisition process do not change scaling exponents of our model
(see Fig. S2 in Supplementary materials).
In the rest of this study we use the first universal network (union of all
KEGG reactions) in our numerical simulations of the model and the second network in our mathematical analysis.
%
%

\begin{figure}[t]
\centerline{\includegraphics[width=0.55\textwidth]{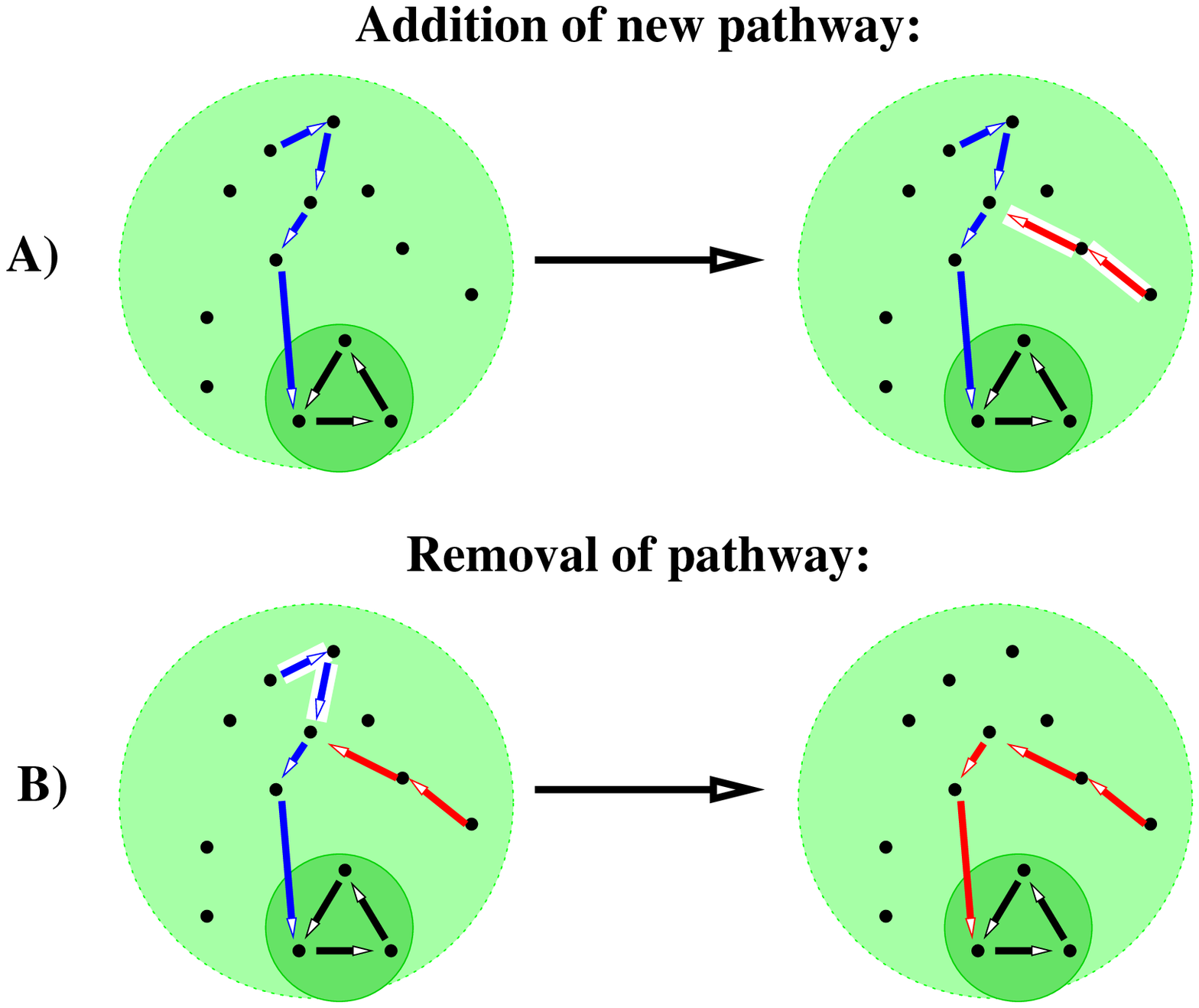}}
\caption{\label{fig1} ``Toolbox'' rules for evolving metabolic
networks in our model.
A) addition of a new metabolic pathway
(red) that is long enough to connect the
red nutrient to a previously existing pathway (blue)
which further converts it to the central metabolic core (dark green).
B) removal of a part of the blue pathway following loss of the blue nutrient.
The upstream portion of the blue pathway that is no longer required
is removed down to the point where it merges with another pathway (red).
The light green circle denotes all metabolites in the universal
biochemistry network from which new pathways are drawn
(see text for details).}
\end{figure}

While toolbox view of evolution is equally applicable to
catabolic (breakdown of nutrients) and
anabolic (synthesis of complex metabolites) pathways,
for simplicity we will simulate only addition and removal
of catabolic branches. Given the repertoire of enzymes of an organism each of the
$N_{univ}$ universal metabolites can be categorized as
either ``metabolizable'' (connected
to biomass production), or
``non-metabolizable'' (currently outside of the
metabolic network).
To add a new branch to the network in our model we first randomly
choose a non-metabolizable molecule as a new nutrient (leaf).
A pathway/branch that begins at the leaf and connects it to the
set of metabolizable molecules is then added to the
network.
This connecting pathway consists of a linear
chain of reactions randomly selected from the universal network
until it {\it first intersects} with
the already existing metabolic network of the
organism. The leaf plus all the
intermediate metabolites of this branch thereby become
metabolizable. This process is illustrated in Fig.
\ref{fig1}A.

In our model
pathway additions and removals are treated in a
symmetric fashion.
The steps leading to pathway deletion
are illustrated in Fig. \ref{fig1}B. First, one of the
leaves of the network corresponding to a vanished
nutrient is chosen randomly. The branch
starting at this nutrient/leaf
is followed downstream to the point where it first
intersects another branch of the network.
This entire path, starting from the leaf down to the
merging point with another pathway is then removed from the network. The
selected nutrient along with all intermediate
metabolites thereby become non-metabolizable.
%

The network in our model evolves by a random sequence of
pathway additions and removals (see Methods for more
details).
%
Since our goal is to
understand how properties of metabolic and regulatory
networks scale with the genome size of an organism, we
take multiple snapshots of the evolving network
with different values of $N_{met}$
-- the current number of nodes in the metabolic network,
which in our model is equal to the number
of reactions or metabolic enzymes.
%
%


\subsection{Assigning transcriptional regulators to
metabolic pathways}

Operation of metabolic networks involves
regulating production of enzymes in response to nutrient
availability. In prokaryotes most of this
regulation is achieved at the transcriptional level.
In order to investigate the interface
between metabolic and regulatory
networks we extend our model to include transcription
factors (TFs) which are activated by nutrient availability
to turn on or off the enzymes in individual metabolic
pathways. In the basic version of our
model shown in Fig. \ref{fig2}A we
chose the following simple method to assign TFs to reactions:
one randomly picks a leaf/nutrient
and follows its reactions
downstream until this branch either reaches the
metabolic core
or merges with a
pathway regulated by a previously assigned TF.
A new TF is then assigned to regulate all
reactions in this part of the nutrient utilization
pathway.
This process is repeated until all
enzymes/reactions have been assigned a (unique)
transcriptional regulator  (see Fig. \ref{fig2}A).
Each TF is activated by the presence of
the corresponding nutrient in the environment.
Note that this method results in exactly one TF per nutrient,
and that the out-degree distribution of TFs in the regulatory
network is identical to the distribution of branch lengths in
the metabolic network.

In addition to this simple regulatory network architecture
we have tried several others illustrated in Figs.
\ref{fig2}B-D. The advantage of these more
complicated schemes is that they ensure that
on/off states of connected metabolic pathways are
properly coordinated with each other.
For example, unlike in Fig. \ref{fig2}A, in Figs.
\ref{fig2}B-D the red transcription factor (TF2)
turns on the downstream (and only the downstream)
part of the blue pathway necessary for utilization of the
red nutrient. We will further compare network topologies
generated by these rules in the Discussion section.

\subsection{Comparison of the
model with empirical data}

In agreement with the toolbox argument outlined in the
introduction, we found (see Fig. \ref{fig4}A) that the number of
transcriptional regulators of an organism
scales steeper
than linearly with the total number of metabolites in its metabolic network,
which in our model is equal to its number of reactions or enzymes:
\begin{equation}
N_{TF}\propto (N_{met})^{\alpha}.
\end{equation}
The best fit has $\alpha=1.8\pm 0.2$.
In Fig. \ref{fig4}A we directly compare numerical simulations of the toolbox
model (red diamonds) to the empirical scaling of the number of transcription factors
with the number of genes in all currently sequenced prokaryotic genomes
(green circles). To approximate the total number of genes $N_{genes}$ in the
whole genome we multiplied the number of metabolites/reactions $N_{met}$
by a constant factor.
The empirical value of the
ratio $N_{met}/N_{genes} \sim 0.2$ was estimated as follows:
metabolic enzymes constitute about a quarter of
all genes in a procaryotic genome independent of its size
(see blue line in Fig. 1a of \cite{nimwegen}).
%
%
Due to
presence of isoenzymes the number of different
reactions  catalyzed by these
enzymes (equal to the number of metabolites $N_{met}$
in our model) is somewhat smaller and its average
value over 451 fully sequenced prokaryotic genomes \cite{metapath}
is 20\%.
The model results shown in Fig. \ref{fig4} were simulated
on the universal network
formed by the union of KEGG reactions in all organisms.
However, a model simulated on a random universal network of the same size
$N_{univ} \simeq 1800$ produced essentially identical results (black crosses in Fig. S1). This agreement indicates that
the scaling between
$N_{TF}$ and $N_{met}$
for the most part is determined by just
the number of universal metabolites -- $N_{univ}$,
and is not particularly sensitive to the topology of connections between them.
On the other hand, we believe that nearly precise agreement of
the actual number of regulators in real prokaryotic genomes and in the model
is coincidental. Indeed, even in prokaryotes not all transcription factors
are dedicated to regulation of metabolic enzymes. This means that to represent
all TFs in the whole genome the number
of metabolic transcription factors in our model has to be multiplied by a
currently unknown number.
Furthermore, as discussed in the beginning of the Results section
the effective size of the universal network for
real-life horizontal transfer of metabolic pathways is likely to be
different from the union of all reactions currently listed in KEGG.
We still believe that the KEGG-based universal network provides
a correct order-of-magnitude estimate of $N_{univ}$. Hence, the
approximate agreement between $N_{TF}$ vs $N_{genes}$ plots in
our model and real prokaryotic genomes is an encouraging sign.

In addition to providing an explanation to
the quadratic scaling between numbers of
leaves and all nodes, our model nicely
reproduces the large-scale topological structure
of real-life metabolic networks.
An example of a metabolic network generated by the toolbox model
is shown in Fig. \ref{fig3}B.
Its tree-like topology reflects
our simplification that each reaction converts a single
substrate to a single product. The network
is hierarchical in the sense that smaller linear pathways
tend to be attached to progressively longer and longer pathways, until they
finally reach the metabolic core. This architecture is
reminiscent of drainage networks in which many
short tributaries merge to give rise to larger rivers.
For comparison, in Fig. \ref{fig3}A we show a tree-like
backbone (to match linear pathways in our model) of the {\it E.
coli} metabolic network \cite{kegg,metapath} of
approximately the same size as the model network
in Fig.  \ref{fig3}B. The details
of generating this backbone are described in the Methods section.
The overall topological structure of real and model networks
clearly resemble each other.

To better quantify this visual comparison in Fig. \ref{fig4}B we compare
cumulative branch length distributions $P(K_{out} \geq K)$ in our model
with $N_{met}=400$ (red diamonds for $N_{univ}=1800$ and red squares for
$N_{univ}=900$)
and in real metabolic network in {\it E. coli} of comparable size (green circles).
All three distributions are characterized by a long
powerlaw tail:
$P(K_{out}) \sim K_{out}^{-\gamma}$.
Best fit value of the exponent
$\gamma=2.9 \pm 0.2$ is similar in model and real-life
networks and agrees with our analytical result $\gamma=3$ derived in
the next section.
Furthermore, the data in our model
simulated on a truncated universal network
with $N_{univ}=900$ (red crosses in Fig. \ref{fig4}B calculated for
the red network in Fig. \ref{fig3}B) are in
excellent agreement with their real-life conuterpart in
{\it E. coli} (green circles in Fig. \ref{fig4}B calculated for
the green network in Fig. \ref{fig3}A)
throughout the whole range.

%

In Fig. S3 we compare distributions of regulon sizes (branch lengths) in our model (red diamonds in Fig. \ref{fig4}B) and in
the Regulon database
\cite{regulon} including all presently known transcriptional regulations in
{\it E. coli}. One can immediately see that the tail of the
distribution in the Regulon database with the exponent $\simeq 2$
is considerably broader than in our model. There are several possible explanations
of this discrepancy: 1) coordination of activity of different metabolic pathways with each other  as shown in Figs. \ref{fig2}B-D inevitably increases out-degree of transcription factors and gives rise to larger regulatory hubs;
2) regulation of proteins other than metabolic enzymes
in the same regulon; 3) an anthropogenic effect in which
better studied transcription factors included in the regulon database
have larger-than-average out-degrees.
In the Discussion section we return to comparison
real-life and model regulatory networks in more details.

\subsection{Mathematical derivation of scaling behavior in toolbox model}

When a new nutrient (leaf) is added to a network of
size $N_{met}$,
the length of the metabolic pathway
required for its utilization is (on average)
inversely proportional to $N_{met}$. This result is easy to show
for a mean-field version of the model on a randomly generated
universal network.
In this case each reaction in the new pathway
has the same probability $p=N_{met}/ N_{univ}$ to end in one
of the $N_{met}$ currently metabolizable molecules.
The minimal pathway required for utilization of the new nutrient
involves only the reactions
until the first intersection with the already existing metabolic network.
The average length of such pathway is just the inverse of this
probability: $1/p=N_{univ}/N_{met}$.
When this pathway is added, the number of
metabolizable molecules increases by
$\Delta N_{met}=N_{univ}/N_{met}$ and the number of regulators
increases by one: $\Delta N_{TF}=1$. In the steady state of the model,
removal of a branch produces the opposite result:
$\Delta N_{met}=-N_{univ}/N_{met},~~ \Delta N_{TF}=-1$.
In both cases one has:
\begin{equation}
\frac{dN_{met}}{dN_{TF}} \; = \; \frac{N_{univ}}{N_{met}}
\label{eq-growth}
\end{equation}
the integration of which gives
\begin{equation}
N_{TF} =\frac{N_{met}^2}{2N_{univ}}.
\end{equation}
Therefore, the quadratic scaling
between $N_{TF}$ and $N_{met}$ naturally emerges from our toolbox
model.

Similar calculations allow one to derive the
scale-free distribution of branch lengths (regulon sizes)
in our model:
\begin{equation}
N(K_{out}) \sim K_{out}^{-\gamma}, \quad \mathrm{with } \quad \gamma=3.
\label{eq:gamma}
\end{equation}
Indeed, the expected length of a newly
added metabolic pathway
(or the out-degree of its regulator in transcription regulatory network
shown in Fig. \ref{fig2}A) is $K_{out}=N_{univ}/N_{met}$.
As the size of the metabolic
network increases, the length of each new pathway
progressively shrinks.
If the network was monotonically growing,
longer pathways of length $K_{out} \ge K$ were added at the
time when the number of metabolites was smaller than
$N_{univ}/K$ or equivalently the
number of transcription factors was below
$N_{univ}/(2 K^2)$. Therefore,
$P(K_{out} \ge K)=N_{univ}/(2 K^2)/N_{TF}$ or
$P(K_{out}=K) \sim N_{univ}/(N_{TF} K^3)$ so that
$\gamma=3$ in Eq. \ref{eq:gamma}.
As evident from Fig. \ref{fig4}B, random cycling through
addition and removal of pathways in the steady state of our model
does not significantly change this exponent with best fit value
of $\gamma=2.9 \pm 0.2$ shown as solid line in Fig. \ref{fig4}B.

\section{Discussion}

\subsection{Trends of average in- and out-degrees in
the regulatory network as a function of genome size}

As was pointed out by van Nimwegen
\cite{nimwegen,van_nimwegen_book,van_nimwegen_2008}
faster-than-linear scaling of the number of
transcription factors generates systematic differences in topology of
transcriptional regulatory networks as a function of genome size.
Indeed, the total number of regulatory interactions
(pairs of TFs and their target genes)
in a given organism can be written either as
$N_{genes} \langle K_{in} \rangle$ if one counts the
incoming regulatory inputs of all genes, or as
$N_{TF} \langle K_{out} \rangle$ if one counts the
regulatory outputs of all transcription factors.
Here the brackets
denote the average over a given genome. Therefore,
one always has
\begin{equation}
\frac{N_{TF}}{N_{genes}}=\frac{\langle K_{in}
\rangle}{\langle K_{out} \rangle} \label{degree_ratio}
\end{equation}
The empirical data \cite{stover,nimwegen} indicate
that the left hand side of this equation monotonically
grows with genome size and is roughly proportional to
$N_{genes}$. Therefore, an increase in
the number of genes in larger genomes
must be accompanied either by an increase in average
in-degree $\langle K_{in} \rangle$ of all genes or by
a decrease in average out-degree $\langle K_{out}
\rangle$ of transcriptional regulators.
The latter trend is indirectly supported by
the empirical observation  \cite{van_nimwegen_book} that
the average operon size (a lower bound on regulon size)
is negatively correlated with $N_{genes}$.
This trend also exists in  our basic model (Fig. \ref{fig2}A)
in which $K_{out}$ of transcription factors
regulating newly added metabolic pathways progressively
decreases with $N_{met} \sim N_{genes}$.
Furthermore, another recent study
\cite{van_nimwegen_2008} found
no systematic correlation between $\langle K_{in} \rangle$ and $N_{genes}$.
%
This is the case in our model in Fig. \ref{fig2}A where
all enzymes representing vast majority of all proteins in our model
have the same $K_{in}=1$ independently of genome size.
However, such complete lack of coordination between different metabolic
pathways is not realistic. To correct this
we explored several other regulatory network architectures illustrated in
Figs. \ref{fig2}B-D. In all these models enzymes are regulated by more than one transcription factor. Transcription factors in the
model in Fig. \ref{fig2}B ensure a
complete top-to-bottom regulation of the entire
pathway for utilization of each nutrient.
In this case centrally positioned
metabolites have unrealistically large in-degrees.
Opposite to the basic model
in Fig. \ref{fig2}A, the average in-degree
$\langle K_{in} \rangle$ in Fig. \ref{fig2}B increases with $N_{genes}$,
while $\langle K_{out} \rangle$ remains constant. Real-life regulatory
networks are likely to be somewhere in-between these two extreme scenarios
illustrated in Figs. \ref{fig2}A \ref{fig2}B.

\subsection{Coordination of activity of upstream and downstream
metabolic pathways}

\begin{figure}[t]
\centerline{\includegraphics[width=0.55\textwidth]{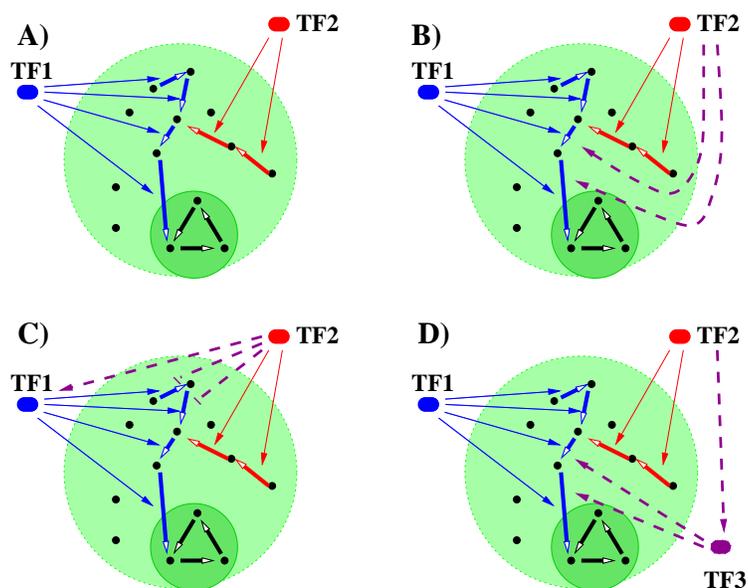}}
\caption{\label{fig2} Schematic diagrams illustrating several possible regulatory
network architectures for control of metabolic enzymes/pathways.
Four panels correspond to different versions of our model discussed
in the text. In the basic model (panel A) there is no coordination
of activity between red and blue metabolic pathways. More
realistic models (panels B-D)
include extra regulatory interactions (purple dashed lines)
and transcription factors (purple TF3 in panel D) ensuring
that only the part of the blue pathway necessary for utilization
of the red nutrient is turned on by the corresponding
transcription factor (red TF2).
}
\end{figure}

\begin{figure}[t]
\centerline{\includegraphics[width=0.5\textwidth]{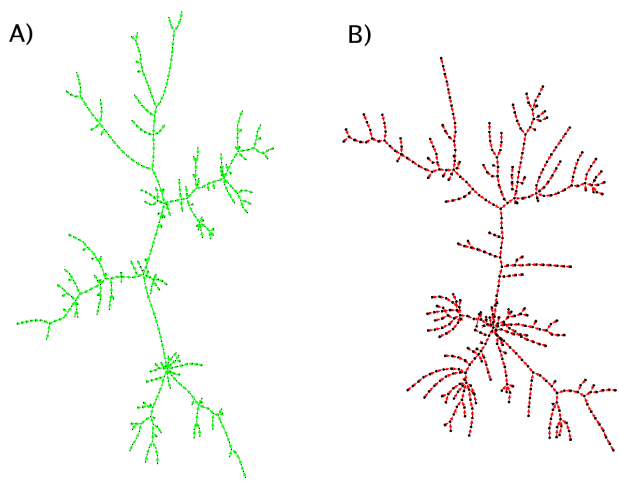}}
\caption{\label{fig3}
A. The backbone of the metabolic network in {\it E. coli} \cite{kegg}
located upstream of central metabolism (green).
B. A similarly-sized network generated by
our model (red).
Note hierarchy of branch lengths in both panels in which shorter
pathways tend to be attached to
progressively longer pathways.
The branch length distributions in real and model networks
are shown as green circles and red squares in Fig. \ref{fig4}B.
}
\end{figure}

\begin{figure}[t]
\centerline{\includegraphics[width=0.7\textwidth]{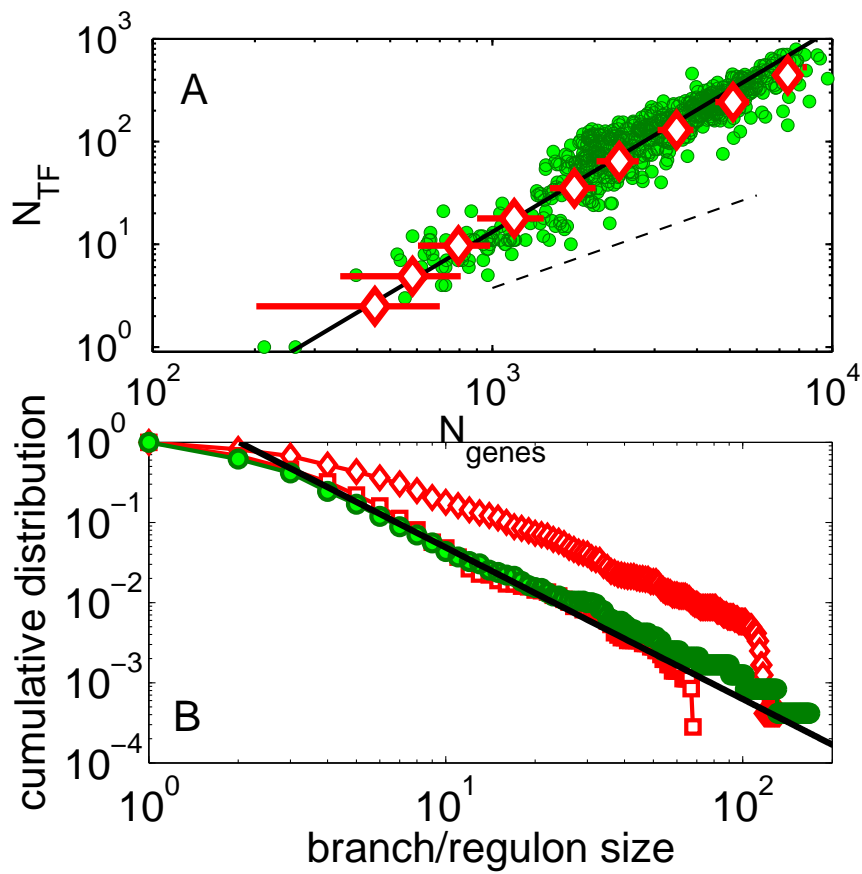}}
\caption{\label{fig4} A. The number of transcription
factors scales approximately quadratically with the total number of
genes in real prokaryotic genomes \cite{kegg,DBD} (green)
and our model (red) simulated on the KEGG
universal network with $N_{univ}=1800$. The number
of metabolic reactions in the model was rescaled to
approximate the total number of genes in a genome (see text for more details).
Error bars correspond to data scatter in multiple simulations of the model.
The solid line with slope 2 is the best
powerlaw fit to the scaling in real prokaryotic genomes
(the best fit to our model is $1.8 \pm 0.2$), while the dashed line with slope
1 is shown for comparison to emphasize deviations from linearity.
B. Cumulative distributions of pathway/branch lengths
in the {\it E. coli} metabolic network (green circles)
and our model of comparable size (red symbols) have similar
tail exponents. The negative slope of the
best powerlaw fit $\gamma-1=1.9 \pm 0.2$ (solid line)
is consistent with our analytical result $\gamma=3$
(see text for details).
The toolbox model with $N_{met}=400$ was
simulated on universal networks of KEGG reactions
with $N_{univ}=1800$ (red diamonds) and
$N_{univ}=900$ (red squares) nodes.
}
\end{figure}

Converting a nutrient into biomass of an organism often
involves several successive pathways each regulated by
its own transcription factor. States of activity of
such pathways have to be coordinated with each other.
Our basic model illustrated in Fig. \ref{fig2}A
does not involve such coordination. In this model:
\begin{itemize}
\item
Transcription factors do not regulate
other transcription factors.
This results in ``shallow''
transcriptional regulatory networks consisting of only
two hierarchical layers: the upper level including all
regulators, and the lower level including
all workhorse proteins (metabolic enzymes). While this
assumption in its pure form is certainly unrealistic,
it approximates the hierarchical structure of real
prokaryotic regulatory networks, which were
shown to be relatively shallow
\cite{thieffry,shen-orr,isambert_pnas2007}. That is to
say, the number of hierarchical layers in these
networks was shown to be smaller than expected by pure
chance \cite{isambert_pnas2007}.
\item
In the regulatory network shown in Fig. \ref{fig2}A
every enzyme is regulated by precisely one
transcription factor. Once again this feature, while obviously
unrealistic, approximates topological
properties of real-life regulatory
networks e.g one in {\it E. coli}.
In \cite{thieffry} it was shown that in this network
the in-degree distribution peaks at one regulatory
input per protein beyond which it rapidly
(exponentially) decays. This should be contrasted with
a broad out-degree (regulon size) distribution
\cite{thieffry} which has a
long power-law tail reaching as high as
hundreds of targets.
\end{itemize}

Several possible regulatory network architectures
ensuring necessary coordination of activity of
upstream and downstream pathways are shown
in Figs. \ref{fig2}B-D.
Models shown in Fig. \ref{fig2}C-D solve the coordination
problem by adding regulatory interactions among transcription
factors.
The positive regulation TF2 $\to$ TF1 in Fig. \ref{fig2}C ensures
that the nutrient processed by the red pathway would
be converted to the central metabolism (dark green area) by the
downstream part of the blue pathway\footnote{Note that in biosynthetic
(anabolic) pathways the direction of metabolic flow is opposite to that
in a nutrient-utilization (catabolic) pathways used in our illustrations (Fig. \ref{fig2}A-D).
As a result, the direction of regulatory interactions between transcription factors
should be reversed as well. Thus in biosynthetic pathways
one expects more centrally-positioned regulator with larger out-degree
to regulate its more peripheral (and less connected) counterparts as is
known to be the case e.g. in the leucine biosynthetic pathway
(see \cite{haoli2008} and references therein)}.
One problem with adding the TF2 $\to$ TF1 regulation
is that it stimulates some unnecessary enzyme production. Indeed,
the presence of the red nutrient triggers the production of enzymes of the entire blue pathway including those located upstream of the merging point with the red pathway
which are not required for red nutrient utilization.
To eliminate this waste of resources we added negative
regulations of these upstream enzymes by TF2 (see Fig. \ref{fig2}C).
Other architectures shown in Fig. \ref{fig2}B and \ref{fig2}D instead of suppressing the upstream enzymes of the blue pathway exclusively activate
its downstream enzymes.
In Fig. \ref{fig2}B transcription factors
regulate the entire length of the long path from
every leaf (nutrient) all the way down to central metabolism.
Another option illustrated in Fig. \ref{fig2}D
is to add a new transcription factor (TF3) activated by the TF2
to regulate only the downstream part of the blue pathway.
Even though the number of transcription factors in Fig. \ref{fig2}D
is up to two times larger than the number of leaves in the
metabolic network, we have verified that their quadratic scaling
remains unchanged.

Transcription regulatory networks are
also characterized by a large number of
feed-forward loops
\cite{shen-orr}.
It has been also
conjectured \cite{shen-orr}
that some of them serve as low-pass filters
buffering against transient
fluctuations in nutrient availability.
Such loops could be
easily incorporated in our models.
One possibility would be to add regulatory interaction between TF2 and TF1 in
Fig. \ref{fig2}B. For the model in Fig. \ref{fig2}D one might
extend the range of TF2 to include at least part of the targets of TF3
and/or add a regulatory interaction between TF1 and TF2.
%
%
Our simulations of models in Fig. \ref{fig2}B-D
indicate that they all give rise to very long regulons. The distribution of regulon
sizes of these models shown in Fig. S4 has a tail significantly
broader than the one empirically observed in {\it E. coli} \cite{regulon}.
A detailed study of regulatory network architectures used by
real-life prokaryotes to ensure coordination of activity
of their metabolic pathways goes beyond the scope of this study
and will be addressed in our future research.

\subsection{Prokaryotic genomes are shaped by horizontal gene transfer and prompt removal of redundant genes}
The Horizontal Gene Transfer (HGT) of whole modules of
functionally related genes from other organisms is the
likely
mechanism by which new pathways are added to
the metabolic network in our model. Indeed,
the rules of our model imply that an organism acquires
several enzymes necessary to utilize a new nutrient not
one by one but all together.
Indeed, a pathway converting a nutrient to
a downstream product that is disconnected from the rest of
the metabolic network does not contribute to biomass production
and thus confers no evolutionary advantage to the organism.
The dominant role of HGT in shaping contents of
prokaryotic genomes in general and their metabolic networks in particular
is well documented \cite{koonin_review}. For example,
a recent empirical study \cite{horiz_transfer}
reports that
horizontally
transferred enzymes
\begin{itemize}
\item Outnumber duplicated enzymes during the last 100
million years in evolution of {\it E. coli}.
\item Frequently confer
condition-specific advantages, facilitating adaptation to new environments.
As a consequence, horizontally-transferred pathways tend
to be located at the periphery of the metabolic
network rather than near its core.
\item tend to come in functionally-coupled groups (see also \cite{gelfand_mirny2006} for
a genome-wide analysis of this trend).
\end{itemize}
These empirical observations make the
central assumptions of our model all the more plausible.
Another feature of evolution of prokaryotic genomes used in
our model is their tendency to promptly
remove redundant genes.
Indeed, in our model we implicitly assume that if a set
of horizontally transferred genes contains
some enzymes that are already encoded in the
genome, these redundant copies are promptly removed.
Stopping the added metabolic branch precisely at the intersection
point with the existing metabolic network
corresponds to instantaneous removal of these redundant genes.
We verified that this simplification could be relaxed
without changing scaling exponents of the model .
This is demonstrated in Fig. S2 in supplementary materials
where we simulated a version of our model assuming more realistic
finite rate of removal of redundant genes.

Both these features (massive horizontal gene transfers
and prompt removal of redundant genes)
are not characteristic of eukaryotic genomes in general,
and those of multicellular organisms in particular.
That is consistent with our finding of approximately
linear scaling of $N_{TF}$ with $N_{genes}$ in
genomes of animals (see Fig. S5 where the best fit exponent $1.15 \pm 0.2$).
The best fit exponent for all eukaryotic genomes ($1.3 \pm 0.2$ \cite{nimwegen})
is marginally higher and is still much lower than its value in prokaryotes
($2.0 \pm 0.2$).

Several earlier modeling efforts \cite{nimwegen,kauffman,enemark}
explained the quadratic scaling
in terms of gene duplications followed by divergence of the
resulting paralogs.
Models of this type assume that additions and deletions of
individual genes are
to a large degree decoupled from their biological function.
Conversely, our model is, to the best of our knowledge, the
first attempt to explain this scaling relation
in purely functional terms. Instead of single genes we add and delete
larger functional units (metabolic pathways) and assume that they
are retained by evolution only if they positively contribute to the
functioning of the organism, that is to say if they get
connected to its biomass production through the existing
metabolic network. Also, contrary to earlier explanations
\cite{nimwegen,kauffman,enemark}, our toolbox model relies
on a different evolutionary mechanism (horizontal
gene transfer vs gene duplications) that is predominant in
prokaryotes.

\subsection{How quickly do new pathways acquire transcriptional regulators?}
In our model we assume that the regulatory network
closely follows changes in the metabolic toolbox of the organism.
For the sake of convenience in our simulations
we choose to assign regulators {\it de novo}
to each new state of the metabolic network.
To verify that this simplification does not distort
our final results we studied a variant of our model in
which the transcriptional regulatory network dynamically
follows changes in the metabolic network.
The regulon size distribution in this model
was essentially unchanged from
the case where regulators
were assigned {\it de novo}.
Such nearly immediate assignment of regulators to newly acquired pathways
is supported by the empirical study of Price and
collaborators \cite{horiz_transfer_reg_fast} reporting
that horizontally transferred
peripheral metabolic pathways frequently include their own transcriptional regulators.
This should come as no surprise, given many well known cases where metabolic enzymes
and their regulators either belong to the same operon or are located
very close to each other on the chromosome
(as e.g. the Lac repressor and the Lac operon).
Our model is also consistent with the selfish operon theory
\cite{selfish_operon} stating that genomic proximity of functionally
related genes is favored by evolution
since it increases the likelihood of a successful
horizontal transfer of a fully functional pathway.

Overall, the emerging consensus
\cite{gelfand_review2006} is that regulatory networks in prokaryotic genomes are
flexible, quickly adaptable, and rapidly divergent
even between closely related strains.
Thus, even in cases when a horizontally transferred pathway does not include a
dedicated transcriptional regulator it could nevertheless be quickly
acquired in a separate HGT event or created by gene duplication of
another TF in the genome.

%

%
%

\section{Materials and Methods}
\subsection{Numerical simulations of the model}
Metabolic network in our model is shaped by
randomly repeating pathway addition and pathway removal steps.
The boundary conditions for this stochastic process do not allow the
number of metabolites to fall below 40 or exceed about 1600.
Networks with different values of $N_{met}$ are then sampled and analyzed.
The universal network used in our study consists of the union of
all reactions listed in the KEGG database \cite{kegg}.
The directionality of reactions and connected pairs of metabolites
are inferred from the map version of the reaction formula:
\url{ftp://ftp.genome.jp/pub/kegg/ligand/}
\url{reaction/reaction_mapformula.lst}.
Since our goal is to model the conversion of nutrients to organism's biomass we kept
the metabolites located upstream of the central metabolism (reachable by a directed path from Pyruvate). This left us with 1813 metabolites connected by 2745 edges.
The exact size and topological structure of the universal network is not known. To test our model on a universal network of a different size (red squares in Fig. \ref{fig4}B)
we pruned the KEGG network down to $\sim 900$ metabolites. This pruning was
achieved by randomly removing nodes along with branches that got disconnected
from the central metabolism. In yet another version shown in Supplementary
Fig. S1 the universal network is
made of random walks on the fully connected graph of $N_{univ}=1800$ metabolites.
From this figure it follows that properties of our model are mainly determined by
the number of nodes in the universal network and not by details of its topology.

%

1) Pathway addition.
One randomly chooses a new leaf (nutrient)
and a self-avoiding random
walk on the universal network.
This directed walk is started at the leaf
and extended until it first intersects the
subset of $N_{met}$ presently metabolizable
molecules. The leaf plus all the intermediate metabolites
of this new branch thereby become metabolizable.

2) Pathway deletion.
One of the $N_{TF}$ network leaves (nutrients) is chosen randomly.
The links downstream from this leaf are followed until the first
merging point of two metabolic branches. All the metabolites down
to this merging point are removed from the network,
thereby becoming non-metabolizable.

We typically choose to begin all simulations with 20
nodes in the ``metabolic core'' (the dark green central circle
in Figs. \ref{fig1}-\ref{fig2}) that are already metabolizable.
This core could be thought of as the ``universal central
metabolism'' present in most organisms.
The number of these core
metabolites, $N_{core}$, is the second parameter of
our model. However, in practice, as long as $N_{core}\ll
N_{univ}$, the network topological structure in the
steady state does not depend on the value of
$N_{core}$. In our simulations we also tried different
starting sets of metabolizable molecules connected by
linear branches to the core but inevitably arrived to
the statistically identical steady-state networks.

\subsection{Sources of empirical datasets}


The distribution of branch lengths
in Fig. \ref{fig2}A
was calculated as follows: first a leaf was randomly chosen
and followed to the metabolic core. Subsequent branches were followed until the
merging point with another branch that was previously selected.
In the metabolic network of the K-12 strain of {\it E. coli}
leaves were defined as either 1) having zero in-degree (no production within
the organism) or 2) having an undirected degree of one
(endpoints of linear branches formed by reversible reactions).
The backbone of the {\it E. coli} network was
defined by following random linear paths starting at these leaves and ending
at the intersection with each other or at the metabolic core.
This left us with the network in Fig. \ref{fig3}A of $\sim 420$ metabolite nodes
(including 112 leaves) located upstream of the central metabolism \cite{kegg}.

To estimate the number of transcription factors in different genomes
shown in Fig. \ref{fig4}A (green symbols) we used the
DBD database \cite{DBD} (\url{www.transcriptionfactor.org})
with its manually curated list of 147
Pfam families of transcription factors. The resulting values of
$N_{TF}$ are in good agreement with those obtained in earlier studies
\cite{stover,nimwegen,lorenzo,tiedje}.


\section{Acknowledgments}
Work at Brookhaven National Laboratory was carried out
under Contract No. DE-AC02-98CH10886, Division of
Material Science, U.S. Department of Energy. Work
at Niels Bohr Institute was funded by the Danish
National Research Foundation through the Center
for Models of Life. SK and KS thank the Theory Institute
for Strongly Correlated and Complex Systems at BNL for
the hospitality and financial support during visits where
some of this work was accomplished. We thank Eugene Koonin,
Yuri Wolf, and Mikhail Gelfand for helpful discussions and
critical comments on this manuscript.

\end{article}
%
%
%

%
%
\end{document}